\documentclass[12pt]{article}
\usepackage{latexsym,amssymb,amsmath,slashed}

\usepackage[dvips]{graphicx}
\usepackage{comment}
\usepackage{color}
\usepackage{hyperref}

\newcommand{\bbox}{\lower.2ex\hbox{$\Box$}}

\csname @addtoreset\endcsname{equation}{section}
\usepackage{ifpdf}
\ifx\pdfoutput\undefined
   \pdffalse
   \usepackage{cite}
 \else
   \pdfoutput=1
   \pdftrue
\newcommand{\rf}[1]{(\ref{#1})}
\usepackage{color}

\def\be{\begin{equation}}
\def\ee{\end{equation}}
\newcommand{\ba}{\begin{eqnarray}}
\newcommand{\ea}{\end{eqnarray}}

\renewcommand{\b}{\beta}

\newcommand{\K}{\mathcal{K}}
\newcommand{\B}{\mathcal{B}_\mu}

\newcommand{\cN}{{\cal N}}

\newcommand{\cL}{{\cal L}}

\def\E{{$E_{7(7)}$}}
\def\K{K{\"a}hler}

\begin{document}

\begin{titlepage}
\begin{flushright}
CERN-TH-2016-214\\
\end{flushright}
\vspace{0.5cm}
\hskip 0.5cm
\vskip 0.5cm
\begin{center}
\baselineskip=16pt


{\Large {\bf      Seven-Disk Manifold, $\alpha$-attractors and B-modes}}

\

\

\

 { \large  \bf Sergio Ferrara$^{1,2,3}$},  {\large  \bf Renata Kallosh$^{4}$}

\vskip 0.8cm
{\small\sl\noindent
$^1$ Theoretical Physics Department, CERN CH�1211 Geneva 23, Switzerland\\\smallskip
$^2$ INFN - Laboratori Nazionali di Frascati Via Enrico Fermi 40, I-00044 Frascati, Italy\\\smallskip
$^3$ Department of Physics \& Astronomy and Mani L.Bhaumik Institute for Theoretical Physics, \\\smallskip U.C.L.A., Los Angeles CA 90095-1547, USA\\\smallskip
$^4$ SITP and Department of Physics, Stanford University, Stanford, California
94305 USA 
}


\vskip 2cm

{\bf Abstract}

\end{center}

{\small  Cosmological  $\alpha$-attractor models in $\cN=1$ supergravity are based on hyperbolic geometry of a Poincar\'e disk with the radius square ${\cal R}^2=3\alpha$. The predictions for the B-modes, $r\approx 3\alpha {4\over N^2}$,  depend on moduli space geometry and are robust for a rather general class of potentials.
 Here we notice that starting with M-theory compactified on a 7-manifold with $G_2$ holonomy, with a special choice of Betti numbers, one can obtain d=4 $\cN=1$ supergravity with rank 7 scalar coset $\Big[{SL(2)\over SO(2)}\Big]^7$. In a model where these 7 unit size 
Poincar\'e disks have identified moduli one finds that $3\alpha =7$. Assuming that the moduli space geometry of the phenomenological models is inherited from this version of M-theory, one would predict  $r \approx  10^{-2}$ for  $N=53$ e-foldings. We also describe the related maximal supergravity and M/string theory models leading to preferred values
$3\alpha =1,2,3,4,5,6,7$.

}\vspace{2mm} \vfill \hrule width 3.cm \vspace{1mm}
{\footnotesize \noindent Sergio.Ferrara@cern.ch, kallosh@stanford.edu }
\end{titlepage}
\addtocounter{page}{1}

\bibliography{supergravity}
\bibliographystyle{toinemcite}

\section{Introduction}

To compare the predictions of theoretical models with the observational  data on inflationary cosmology \cite{Planck:2015xua} one has to use some form of  d=4 Einstein theory.  In particular one can use $\cN=1$ supergravity models making a choice of the \K\, potential and a superpotential to fit the data. Cosmological models called $\alpha$-attractor models \cite{Kallosh:2013yoa,Carrasco:2015uma,Ferrara:2015tyn}, based on hyperbolic geometry of a Poincar\'e disk with the radius square $3\alpha$, are in good agreement with the data.  The tilt of the spectrum of fluctuations and  the level of B-models  depend on the number of e-foldings $N$ and on the moduli space curvature ${\cal R}_{\rm K\ddot{a}hler}= -{2\over 3 \alpha}$:
\be
n_{s} \approx 1-{2\over N}\, ,  \qquad r \approx  3 \alpha \, {4\over N^{2}}\, .
\label{data}\ee

This prediction is valid for $\alpha$-attractor models  with $\alpha \lesssim O(10)$  for rather general class of potentials described in 
\cite{Kallosh:2013yoa,Carrasco:2015uma,Ferrara:2015tyn}.
The early versions of these models were derived in \cite{Kallosh:2013yoa}, the more advanced versions  were presented in \cite{Carrasco:2015uma,Ferrara:2015tyn}. At the level of phenomenological $\cN=1$ supergravity any value of $0< 3\alpha< \infty$ is acceptable, so one can view the future detection of the B-modes, or a new bound on $r$, as an experimental information about the curvature of the moduli space  in these phenomenological models.

However,  one may try to motivate certain preferred values of the Poincar\'e disk radius square $3\alpha$ as originating from a  fundamental theory underlying $\cN=1$ supergravity. It was already suggested in \cite{Carrasco:2015uma} that the lowest possible value  $3\alpha=1$, with one unit size Poincar\'e disk, is motivated by a  maximal superconformal $\cN=4$ theory  \cite{Bergshoeff:1980is} and $\cN=4$ pure supergravity without matter \cite{Cremmer:1977tt}. 

In this note we will study the possible origin of the moduli space geometries in maximal $\cN=8$ supergravity and M/string theory. We assume that when the maximally supersymmetric theories are reduced to $\cN=1$ phenomenological $\alpha$-attractor models, some mechanism of generating the required potentials will take place, but the moduli space geometry will be inherited from the more fundamental theories.

In this setting we will find a  reasonably well motivated models of the Poincar\'e disk with radius square $3\alpha$ taking values $1,2,3,4,5,6,7$. In particular, the case with the highest value of $3\alpha=7$ suggests that  $r $ is only slightly below  $10^{-2}$.

Joint analysis of the data from BICEP2/Keck and Planck experiments \cite{Planck:2015xua} yields an upper limit on B-modes, 
 $r\leq 7\times 10^{-2}$.  The new interesting target with preferred values of $\alpha$ originating in  M/string theory, for the number of e-foldings $47<N<57$,  is now
\be
3\alpha = 7: \qquad r \approx  7 \, {4\over N^{2}},  \qquad      0.86  \times 10^{-2} < r < 1.3 \times 10^{-2} \ ,
\label{7d}\ee
and  the lowest one in the context of maximal $\cN=4$ superconformal theory  is
\be
3\alpha = 1: \qquad r \approx   \, {4\over N^{2}}, \qquad     1.2  \times 10^{-3} < r < 1.8 \times 10^{-3}\, .
\label{1d}\ee

\section{Poincar\'e disk with the radius square $3\alpha$}
Consider the \K\, potential \be
K= -3\alpha \ln(1-Z\bar Z) \ .
\label{disk}\ee
It describes a  Poincar\'e disk with the radius square $3\alpha$.
 The metric of the moduli space is 
 $ g_{Z\bar Z}= K_{Z\bar Z}= {3\alpha\over (1- Z\bar Z)^2}
$.
 The \K\, manifold curvature computed from this metric depends on $\alpha$:
\be
 {\cal R}_{\rm K\ddot{a}hler}= - g_{Z\bar Z}^{-1} \partial_Z \partial_{\bar Z} \log g_{Z\bar Z}= -{2\over 3 \alpha} \ .
\label{curv}\ee
The kinetic term for the complex scalar field is
\be
  ds^2=  {3\alpha\over (1- Z\bar Z)^2} d Z d \bar Z = {d x^2+d y^2 \over \Big (1-{ x^2+ y^2\over 3\alpha}\Big )^{2}}    \, , \label{old Lag-alpha}
\ee
where $Z= ( x+i y)/\sqrt {3\alpha}$.
For the vanishing sinflaton\footnote{The sinflaton field  $Z-\bar Z$ can be either stabilized to become heavy, or in models with constrained orthogonal superfields this field depends on fermions and does not participate in cosmological evolution.} 
 the kinetic term becomes in terms of the inflaton $Z= \bar Z= \tanh {\varphi\over \sqrt{6\alpha}}$ 
\be
 3\alpha\, {\partial_\mu Z \partial^\mu  Z \over (1- Z)^2}= {1\over 2} (\partial_\mu \varphi)^2 \ .
\ee
In these models the potentials depend on a geometric variable $Z=\bar Z$
\be
V= V\big (\tanh {\varphi\over \sqrt{6\alpha}}\big)\ .
\ee 
\subsection{Half-plane variables}
One can use an alternative description of the same physical system by making a choice $ {1+Z\over 1-Z}=-i \tau $.
\be
K=  -3\alpha \ln(-i(\tau -\bar \tau)) \ .
\label{HP}\ee
The kinetic term for the complex scalar field is
\be
 ds^2=   3 \alpha {d\tau d \bar \tau \over (2 \, {\rm Im} \tau)^2} \ .
 \ee
In this form the kinetic term has an $SL(2, \mathbb{R})$ symmetry
  \be
\tau'= {a\tau+b\over c\tau +d},  \qquad ad-bc\neq 0 \ ,
\label{sl}\ee
where $a,b,c,d$ are real numbers and 
\be
{ d \tau d \bar \tau\over (\tau-\bar \tau)^2}=  { d \tau' d \bar \tau'\over (\tau'-\bar \tau')^2}\ .     
\ee
When the sinflaton $\tau +\bar \tau$ vanishes
 at $\tau =- \bar \tau = i e^{\sqrt{2\over 3\alpha} \varphi}$
\be
 3 \alpha {d\tau d \bar \tau \over (2 \, {\rm Im} \tau)^2}=  {1\over 2} (\partial_\mu \varphi)^2 \ .
\ee

\section {  Seven-disk geometry  in maximal supergravity}

Before looking at M-theory on a 7-manifold with  $G_2$ holonomy we will explain the origin of the seven-disk geometry starting from D=4 $\cN=8$ supergravity. M theory/d=11 supergravity can be compactified on a 7-torus, which  leads to d=4 maximal $\cN=8$ supergravity  \cite{Cremmer:1979up} upon dualization of the form fields. This model has 70 scalars in the coset space ${E_{7(7)}\over SU(8)}$ and \E\, symmetry. 
Following  \cite{Duff:2006ue}, we consider truncation of $\cN=8$ supergravity \cite{Cremmer:1979up} to $\cN=4$ supergravity interacting with six $\cN=4$ vector multiplets. The \E\,  symmetry is decomposed as follows
\be
E_{7(7)}  \supset [SL(2)] \times SO(6,6) \ , 
\label{N4}\ee
The 70 scalars of $\cN=8$ supergravity \cite{Cremmer:1979up} are first truncated to 
\be
70 \rightarrow 2+36
\ee
In the  next step one takes into account that
\be
SO(6,6) \supset SO(2,2)\times SO(2,2)\times SO(2,2)
\ee
and 
\be
36 \rightarrow 3\times 4
\ee
so that
\be
70 \rightarrow 2(1+6) = 2\times 7=14\, .
\ee
This truncation has a \K\, structure supporting $\cN=1$ supersymmetry. One can
 identify  7 Poincar\'e disks
due to the decomposition 
\be
E_{7(7)} (\mathbb{R}) \supset [SL(2, \mathbb{R})]^7\, .
\label{DF}\ee
 The original kinetic term is reduced to a form with  the \K\, potential of the form
\be
K= - \sum_{i=1}^7   \ln(-i(\tau_i -\bar \tau_i))
\label{7disks}\ee
with 7 pairs of  independent scalars and the $[SL(2, \mathbb{R})]^7$ symmetry, a seven-disk manifold.  The fact that the disk of the $SL(2)$ commuting with $SO(6,6)$ has the same \K\,  curvature of the other six ${SL(2)/SO(2)}$
(each separately corresponding to $\alpha=1/3$) can be understood by string triality arguments \cite{Duff:1995sm} and by the underlying special geometry of the
N=2 truncation \cite{Ferrara:2007pc}.

\section{M theory on a 7-manifold with  $G_2$ holonomy}

 Instead of a compactification on a 7-torus,  one can compactify M theory on a 7-manifold with $G_2$ holonomy. The early investigation of $G_2$ holonomy in physics was performed  in  \cite{Awada:1982pk}, with review of the first 20 years in \cite{Duff:2002rw}. One of the most recent application of this compactification can be found in \cite{Acharya:2016kep} and, of course,  many more studies of M theory on $G_2$ were performed over the years. 
 
 Here we will focus on the model studied in
  \cite{Duff:2010ss,Duff:2010vy}, it requires the following choice of the Betti numbers
\be
(b_0, b_1, b_2, b_3)= (1,0,0,7)\, .
\label{Betti}\ee
This theory  is identified with the maximal rank reduction
on the seven torus and leads directly to
 d=4 $\cN=1$ `curious supergravity' where 7 complex scalars are coordinates of the  coset space
\be
\Big [{SL(2, \mathbb{R})\over SO(2)}\Big ] ^7\, .
\ee
The corresponding \K\, potential describing the scalar sector of this theory is the one in eq. \rf{7disks}
with 7 pairs of  independent scalars and the $[SL(2, \mathbb{R})]^7$ symmetry. This model is one of the  `Four curious supergravities' defined in \cite{Duff:2010vy}. The other 3 have $\cN=2, \,  \cN=4, \, \cN=8$ supersymmetries, we are interested only in $\cN=1$  `curious supergravity'.
It has the field content defined by Betti numbers :
the d=4 fields originating from the d=11 metric $g_{MN}$ are
\ba
&g_{\mu\nu}  &\rightarrow b_0=1\cr
&A_\mu  &\rightarrow b_1=0\cr
&{\cal A} &\rightarrow b_1+b_3=7
\ea
The ones from d=11 gravitino $\psi_M$ are
\ba
&\psi_{\mu}  &\rightarrow b_0+b_1=1\cr
&\chi  &\rightarrow b_2+b_3=7
\ea
The ones from the 3-form $A_{MNP} $ are
\ba
&A_{\mu\nu\rho}  &\rightarrow b_0=1\cr
&A_{\mu \nu} &\rightarrow b_1=0\cr
& A_\mu  &\rightarrow b_2=0\cr
&A  &\rightarrow b_3=7
\ea
To summarize, the field content of the M theory compactified on a 7-manifold with  $G_2$ holonomy and Betti numbers \rf{Betti} is a metric, a gravitino and a 3-form (which has no degrees of freedom, but affects trace anomaly)
\be
g_{\mu\nu}, \, \psi_\mu, A_{\mu\nu\rho} 
\ee
and 7 scalars,  7 spin 1/2  fields and 7  pseudoscalars 
\be
\tau_i= {\cal A}_i +i A_i, \, \, \chi_i \ .  
\ee
The corresponding \K\, geometry is the seven-disk manifold  in \rf{7disks}.

For generic Betti numbers $(b_0, b_1, b_2, b_3)$ these models are known to have a generalized mirror symmetry, which flips one set of Betti numbers into the other one,
\be
(b_0, b_1, b_2, b_3) \quad \rightarrow \quad (b_0, b_1, b_2- \rho/2, b_3+\rho/2)
\ee
and $\rho\equiv 7b_0 - 5b_1 +3b_2 -b_3$  changes the sign.
One of the reason the model we describe here was given a name `curious supergravity' is that it has $\rho=0$, it is a  {\it self-mirror} in the above sense.  It also means that it has a vanishing Weyl anomaly
$
g_{\mu\nu} \langle T^{\mu\nu} \rangle = - {\rho\over 24\times 32 \pi^2} R^{*\mu\nu\rho\sigma} R^{*} _{\mu\nu\rho\sigma}=0, 
$
the presence of the 3-form $A_{\mu\nu\rho}$ is important for this.

To connect this compactified M theory model to $\alpha$-attractor geometry we can make a choice that all moduli in our 7 unit radius disks in \rf{7disks} are identified, namely
\be
3\alpha =7 \, : \qquad \tau_1=\tau_2=\tau_3=\tau_4=\tau_5=\tau_6= \tau_7\equiv  \tau \ .
\label{all7}\ee
 We are left with one Poincar\'e disk of the  radius square  7 times larger than the unit radius square.
   \ba
K=  - \sum_{i=1}^7   \ln(-i(\tau_i -\bar \tau_i))= -   7 \ln(-i(\tau -\bar \tau)) \ .
\label{K7DF}\ea 
  \be
 ds^2=  7 {d\tau d \bar \tau \over (2 \, {\rm Im} \tau)^2} \ .
 \ee
The following interpretation of this identification can be suggested: the diagonal components of the internal space metric $g_{ij}$ are taken to be the same in all 7 directions, $g_{ij}\sim  \delta_{ij} $, and the 3-form $A_{ijk}$, which leads to 7 pseudoscalars in d=4, since $b_3=7$,  is also the same in all directions. An analogous identification was performed in \cite{Witten:1985xb}, where an early dimensional reduction of superstring theories was studied. The resulting d=4 $\cN=1$ supergravity, neglecting the matter fields $C$ in  \cite{Witten:1985xb},  has the following \K\, manifold:
\ba
K=&-&\ln (-i(s-\bar s))- 3\ln (-i(t-\bar t)) \ .
\label{4disks}\ea
We will show in the next section that using string theory compactification on a  product of  3 tori $T_2 \times T_2 \times T_2 \subset T_6 $ one can get the seven-disk geometry.
\ba
K=&-&\ln (-i(s-\bar s))-\ln (-i(t_1-\bar t_1))-\ln (-i(t_2-\bar t_2))-\ln (-i(t_3-\bar t_3))\cr
\cr
&-&\ln (-i(u_1-\bar u_1))-\ln (-i(u_2-\bar u_2))-\ln (-i(u_3-\bar u_3))
\label{7disksString}\ea
Thus, the model \rf{4disks} in  \cite{Witten:1985xb} corresponds to the one in \rf{7disksString} under condition that 
\be
t_1=t_2=t_3=t \, , \qquad u_1=u_2=u_3={\rm const } \ .
\ee
This means that some fields of higher-dimensional geometry were discarded, for example all $u_i$ fields and the difference between $t_i$ fields.
If instead we would impose on \rf{7disksString} the condition
\be
s=t_1=t_2=t_3=u_1=u_2=u_3=\tau
\ee
we would reproduce the \K\, geometry \rf{K7DF} of the single Poincar\'e disk of the  radius square  $3\alpha=7$.  In analogous manner we can get other values
\be
3\alpha = \{ 1,2,3,4,5,6,7\}
\label{seven}\ee
by requiring that
 \ba 
 &3\alpha =7 \, :  &\tau_1=\tau_2=\tau_3=\tau_4=\tau_5=\tau_6= \tau_7 \equiv  \tau  \cr
 &3\alpha =6 \, :  &\tau_1=\tau_2=\tau_3=\tau_4=\tau_5=\tau_6\equiv  \tau  \, ,  \qquad  \tau_7={\rm const }
\label{all 6}\cr
 &3\alpha =5 \, :  &\tau_1=\tau_2=\tau_3=\tau_4=\tau_5 \equiv  \tau  \, ,   \qquad   \tau_6=\tau_7={\rm const }
\label{all 5}\cr
&3\alpha =4 \, :  &\tau_1=\tau_2=\tau_3=\tau_4 \equiv  \tau  \, ,   \qquad   \tau_5=\tau_6=\tau_7={\rm const }
\label{all 4}\cr
&3\alpha =3 \, :  &\tau_1=\tau_2=\tau_3 \equiv  \tau  \, ,   \qquad   \tau_4=\tau_5=\tau_6=\tau_7={\rm const }
\label{all 3}\cr
&3\alpha =2 \, :  &\tau_1=\tau_2 \equiv  \tau  \, ,   \qquad   \tau_3=\tau_4=\tau_5 =\tau_6=\tau_7={\rm const }
\label{all 2}\cr
&3\alpha =1 \, :  &\tau_1 \equiv  \tau  \, ,   \qquad   \tau_2=\tau_3=\tau_4=\tau_5=\tau_6=\tau_7={\rm const }
\label{all 1}
\ea
We illustrate in Fig. 1 the features of  $\alpha$-attractor models \cite{Kallosh:2013yoa,Carrasco:2015uma,Ferrara:2015tyn} with the seven-disk geometry using the recent discussion of B-modes in the CMB-S4 Science Book  \cite{Abazajian:2016yjj}. We show in Fig. 1 predictions of  $\alpha$-attractor models with seven-disk geometry in the $n_s-r$ plane for $N \sim 55$, for the minimal value $3\alpha = 1$ and for the maximal value $3\alpha = 7$. 
\begin{figure}[t!]
\centering
{\includegraphics[width=15cm]{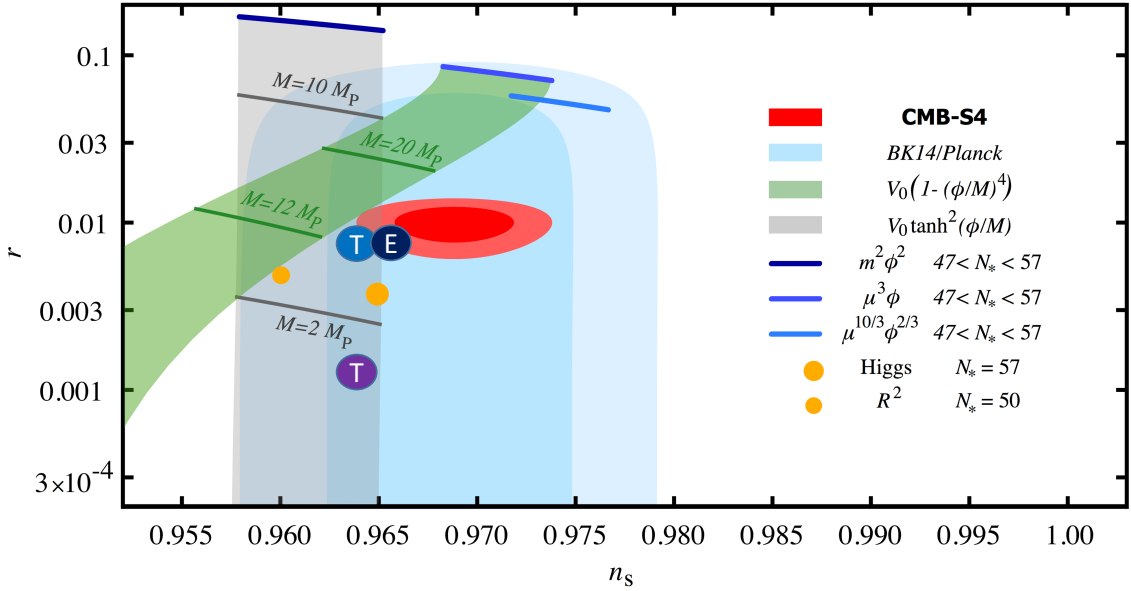}
}~~~
\caption{\label{ff2} \footnotesize This Figure is taken from \cite{Abazajian:2016yjj}, it represents a forecast of CMB-S4 constraints in the $n_s-r$ plane for a fiducial model with r = 0.01. Here the grey band shows predictions of the sub-class of $\alpha$-attractor models \cite{Kallosh:2013yoa,Carrasco:2015uma,Ferrara:2015tyn}. We have  added to this figure a  blue circle with the letter $T$ inside it corresponding to a highest preferred value $3\alpha=7$ 
and the purple one  corresponding to the lowest preferred value $3\alpha=1$  in a seven-disk geometry.  All intermediate cases $3\alpha = \{ 1,2,3,4,5,6,7\}$ are between these two. They all describe the class of $\alpha$-attractor models with $V\sim \tanh^2(\varphi/\sqrt{6\alpha})$,  so-called quadratic $T$-models. The quadratic $E$-models with $V\sim (1-e^{\sqrt{2/3\alpha}\, \varphi})^2$ tend to be slightly to the right of the $T$-models, see \cite{Kallosh:2013yoa}. We show them  as a navy circle with  the letter $E$ inside it. 
}
\end{figure}
\section{Values of  $3\alpha$ in  string theory}
Here we will show how to derive the 7-disk geometry \rf{7disksString} in string theory.
 We start with the  derivation of non-compact symmetries in string theory following \cite{Maharana:1992my}, \cite{Giveon:1994fu}. The toroidal compactification to d=4 of the $\cN=1$ supergravity/string theory in d=10 space-time leads to  scalars in ${SO(6,6) \over SO(6) \times SO(6)}$ coset space\footnote{In general, in the case of the heterotic string theory one finds  scalars in the ${SO(6,6+n) \over SO(6) \times SO(6+n)}$ coset space. Here the scalars in the ${SO(6,6) \over SO(6) \times SO(6)}$ part of the coset space originate from the geometric moduli, whereas the additional ones with $n\neq0$ originate from the matter vector multiplets in d=10. If we  keep some of the vector multiplets, so that $n>0$ we do not find models with Poincar\'e disk geometry.} upon truncation of non-geometric moduli from the d=10 vector multiplets.

 As the result of the  dimensional reduction one finds a d=4 action for the scalars of the following form, 
\be
\int d^4 x \sqrt{-g} \, e^{-\phi} (\cL_1+ \cL_2) \ .
\label{SF}\ee
Here 
\be
\cL_1= R + g^{\mu\nu} \partial_\mu\phi \partial_\nu\phi -{1\over 12} H_{\mu\nu\rho} H^{\mu\nu\rho} \ ,
\label{axiondialton}\ee
and 
\be
\cL_2= {1\over 8} {\rm tr} (\partial_\mu M^{-1} \partial^\mu M ) \ .
\ee
Here $M$ is a symmetric $O(6,6)$ matrix
\be
M=\left(\begin{array}{cc}G^{-1}  & - G^{-1} B  \\B G^{-1} & G- B G^{-1} B\end{array}\right) \ ,
\ee
where $G_{\alpha \beta}$ and $B_{\alpha\beta}$ are the internal space metric and a 2-form, $\alpha, \beta =1,\dots , 6$. Together they represent the 36 coordinates of the coset space ${SO(6,6) \over SO(6) \times SO(6)}$, 
we recover the moduli space of the six-torus $T_6$ in string theory. We would like now to perform the truncation of the 6-torus to three $T_2$ so that
\be
T_2 \times T_2 \times T_2 \subset T_6 
\ee
This corresponds to the reduction $SO(6,6) \supset [SO(2,2)]^3$ and analogous reduction on the on coset representative
\be
{SO(6,6) \over SO(6) \times SO(6)} \rightarrow \Big [{SO(2,2) \over SO(2) \times SO(2)}\Big ]^3 \ .
\ee
This means that we keep the following 9 components of $G_{\alpha \beta}$ 
\be
G_{(IJ)}= (g_{11}, g_{22}, g_{12}; g_{33}, g_{44}, g_{34}; g_{55}, g_{66}, g_{56}) \ ,
\ee
 and 3 components of $B_{\alpha\beta}$
 \be
 B_{[IJ]}=( b_{12}\equiv b_1, b_{34}\equiv b_2, b_{56}\equiv b_3) \ .
 \ee
 We also introduce notation
\be
 g_1\equiv g_{11} g_{22}-g_{12}^2\, ,\qquad  g_2\equiv g_{33} g_{44}-g_{34}^2 \, ,\qquad  g_3\equiv g_{55} g_{66}-g_{56}^2 \ .
 \ee
Now we observe that the coset ${SO(2,2) \over SO(2) \times SO(2)}$ is isomorphic to ${SL(2, \mathbb{R})\over SO(2)} \times {SL(2, \mathbb{R})\over SO(2)}$ and so we can package the  $SO(2,2)$ matrix into an  $SL(2, \mathbb{R}) \times SL(2, \mathbb{R})$  matrix. We will do this for all three copies of ${SO(2,2) \over SO(2) \times SO(2)}$ cosets, following an example of one of them in \cite{Giveon:1994fu}. We have 4 real scalars from $g_{11}, g_{22}, g_{12}, b_{12}$. We package them as follows: $t_1\equiv b_{1} + i \sqrt g_1$ and $u_1\equiv {g_{12}\over g_{22}}+i {\sqrt g_1\over g_{22}}$. The inverse relation is for the 2x2 matrices
\be
G= \left(\begin{array}{cc}g_{11}  & g_{12}  \\g_{21} & g_{22}\end{array}\right)={{\rm Im} \, t_1\over {\rm Im} \, u_1} \left(\begin{array}{cc}u_1u_1^*  & {\rm Re} \, u_1  \\{\rm Re} \, u_1 & 1\end{array}\right)
\ee
\be
B= \left(\begin{array}{cc}0  & b_{12}  \\b_{21} & 0\end{array}\right)= {\rm Re}\,  t_1\left(\begin{array}{cc}0  & 1  \\ -1 & 0\end{array}\right)
\ee
In the same  way we can organize all 6 complex scalars, 3 of them are often called \K\, moduli
\be
t_1= b_{1} + i \sqrt g_1\, , \qquad t_2= b_{2} + i  \sqrt g_2 \, , \qquad t_3= b_{3} + i  \sqrt g_3 \ ,
\ee
and the other 3 are called complex structure moduli
\be
u_1= {g_{12}\over g_{22}}+i {\sqrt g_1\over g_{22}}\, , \qquad u_2= {g_{34}\over g_{44}}+i {\sqrt g_2\over g_{44}} \, , \qquad u_3= {g_{56}\over g_{66}}+i {\sqrt g_3\over g_{66}} \ .
\ee
This corresponds to reorganizing
$
\Big [{SO(2,2) \over SO(2) \times SO(2)}\Big ]^3 
$
into $ \Big [{SL(2, \mathbb{R})]\over SO(2)}\Big ] ^6$. The corresponding \K\, potentials are $K(t_i,\bar t_i) = -\ln (-i(t_i-\bar t_i))$ and $K(u_i,\bar u_i) = -\ln (-i(u_i-\bar u_i))$.

One more important step here is to switch from the string frame as in \rf{SF} to the Einstein frame in d=4, which is a well known procedure of rescaling the metric, see for example \cite{Witten:1985xb}. 
As the result, we find an action with $\cN=1$ supersymmetry and 7 complex scalars. The axion, dual to $H_{\mu\nu\lambda}$,  and dilaton as shown in eq. 
\rf{axiondialton} form a complex scalar
\be
s= a+ie^{\phi}
\ee
with the \K\, potential $K=-\ln (-i(s-\bar s))$.
The complete \K\, potential of the string theory dimensionally reduced on $T_2 \times T_2 \times T_2 \subset T_6 $ is now given by the expression in \rf{7disksString} in the previous section, as promised there.

Thus here again we reproduced the 7 Poincar\'e disk geometry of the unit radius each. We may now study the same cases as we did in the previous section: the conclusion is as in M-theory compactified on 7-manifold with $G_2$ holonomy  in eq. \rf{seven} which gives us seven possible values of $r$, according to \rf{data}, for example for $N=55$
\be
 r\approx \{ 1.3, \, 2.6, \, 3.9, \,  5.2, \,  6.5, \, 7.8, \,  9.1\} \times 10^{-3}
\label{r}\ee

\section{Conclusion}

In conclusion, we made an assumption that the moduli space geometry of the phenomenological $\cN=1$ $\alpha$-attractor models in \cite{Kallosh:2013yoa,Carrasco:2015uma,Ferrara:2015tyn} is inherited from 
the M-theory compactified on 7-manifold with $G_2$ holonomy  to a `curious $\cN=1$ supergravity' \cite{Duff:2010vy},  or from truncated $\cN=8$ maximal supergravity,  or from toroidally compactified string theory. In such case we argued that the possible cosmological  $\alpha$-attractor models might come with the values of $3\alpha=1,2,3,4,5,6,7$ when some of the higher dimensional fields are discarded,  following the procedure employed in the past in \cite{Witten:1985xb} and presented in eq. \rf{all 1}.  To make a step from preferred values for $3\alpha$ to a realistic prediction we would need to find the origin of the suitable class of potentials in these theories.

The relevant preferred values of the ratio of the tensor to scalar fluctuations during inflation are shown in eq. \rf{r}. We illustrated the position of these  models  in $n_s-r$ plane in Fig. 1. The highest one $r\approx 10^{-2}$ will be the first interesting target for the B-mode experiments as well as for the theoretical studies of realistic cosmological models based on the seven-disk geometry.  

\

We are grateful to M. Duff, S. Kachru, A. Linde, A. Marrani, D. Roest, T. Wrase and Y. Yamada for the useful discussions and  to L. Page for explaining to us the importance of relating possible values of $r$ to a  fundamental theory.  The work of SF is supported in part by CERN TH Dept and INFN-CSN4-GSS,
the work of RK is supported by SITP and by the NSF Grant PHY-1316699. We acknowledge the hospitality of the GGI institute in Firenze, where this work was initiated during the workshop `Supergravity: what next?'.

\end{document}

\footnote{ To fit the current data on $n_s$ in \cite{Planck:2015xua} we will use here the number of e-folding of inflation to be $N\approx 55$, so that $n_s\approx 0.9636$. Note, however, that the value of $n_s$  might move with account of the more recent data \cite{Aghanim:2016yuo}.}